# Direct Observation of Morphological and Chemical Changes During the Oxidation of Model Inorganic Ligand-Capped Particles

Maximilian Jaugstetter[1,*], Xiao Qi[2], Emory Chan[2], Miquel B. Salmeron[1], Kevin R. Wilson[3], Slavomír Nemšák[4,5,*], Hendrik Bluhm[6,*]

[1]Materials Sciences Division, Lawrence Berkeley National Laboratory, Berkeley, CA 94720, USA
[2]Molecular Foundry, Lawrence Berkeley National Laboratory, Berkeley, CA 94720, USA
[3]Chemical Sciences Division, Lawrence Berkeley National Laboratory, Berkeley, CA 94720, USA
[4]Advanced Light Source, Lawrence Berkeley National Laboratory, Berkeley, CA 94720, USA
[5]Department of Physics and Astronomy, University of California, Davis, CA 95616, USA
[6]Fritz Haber Institute of the Max Planck Society, D-14195 Berlin, Germany

## ABSTRACT

Functionalization and volatilization are competing reactions during the oxidation of carbonaceous materials and are important processes in many different areas of science and technology. Here we present a combined ambient pressure X-ray photoelectron spectroscopy (APXPS) and grazing incidence X-ray scattering (GIXS) investigation of the oxidation of oleic acid ligands surrounding $NaYF_4$ nanoparticles (NPs) deposited onto $SiO_x/Si$ substrates. While APXPS monitors the evolution of the oxidation products, GIXS provides insight into the morphology of the ligands and particles before and after the oxidation. Our investigation shows that the oxidation of the oleic acid ligands proceeds at $O_2$ partial pressures of below 1 mbar in the presence of X-rays, with the oxidation eventually reaching a steady state in which mainly $CH_x$ and -COOH functional groups are observed. The scattering data reveal that the oxidation and volatilization reaction proceeds preferentially on the side of the particle facing the gas phase, leading to the formation of a chemically and morphologically asymmetric ligand layer. This comprehensive picture of the oxidation process could only be obtained by combining the X-ray scattering and APXPS data. The investigation presented here lays the foundation for further studies of the stability of NP layers in the presence of reactive trace gasses and ionizing radiation, and for other nanoscale systems where chemical and morphological changes happen simultaneously and cannot be understood in isolation.

**TOC FIGURE**

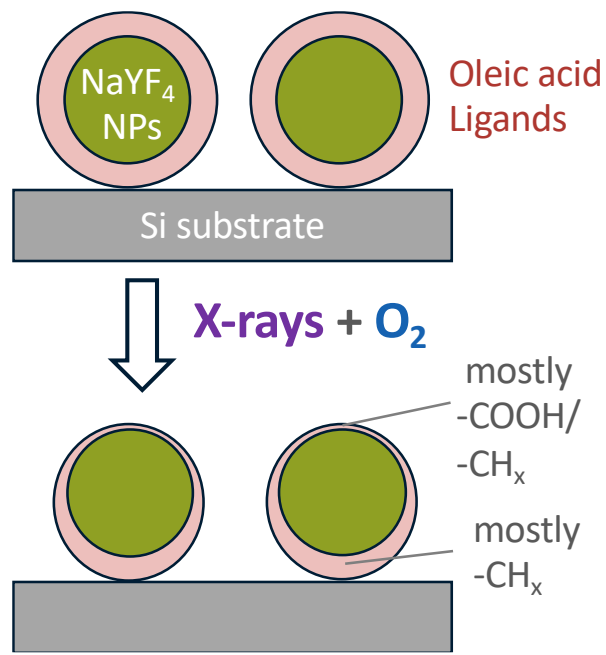

# 1. INTRODUCTION

Organic ligands are widely used as protective layers around nanoparticles (NPs) to prevent the particles from coagulation, to passivate surface charge traps, and to provide chemical stability.[1] The efficacy of the ligand layer depends on the strength of the bonding of the ligand molecules to the NP, the ligand interaction with the surrounding medium (often an organic or aqueous solution), and its stability over time. The protective properties of the capping ligand might considerably change if their molecular structure is altered through chemical reactions. One example is the oxidation of unsaturated fatty acids[2], which are widely used as ligands, by ozone, which is a trace gas in the atmosphere and is also produced through the interaction of ionizing radiation with, for instance, oxygen and water vapor. For the understanding of the stability – or adversely, the aging – of NP-ligand bonds under realistic environmental settings of, e.g., sunlight and aerobic conditions, it is important to investigate both the chemical and morphological changes of the ligands over time.

In the present work we monitor the oxidative degradation of the oleic acid ligand layer surrounding $NaYF_4$ nanoparticles (NPs). When doped with lanthanide ions such as $Tm^{3+}$, $NaYF_4$ nanoparticles can facilitate photon up-conversion[3,4] and photon avalanching[5,6,7], whose applications include microscale lasing[8] and sub-diffraction imaging[5,6]. These NPs can also facilitate down-conversion as scintillators for X-ray[9] and electron imaging[10]. Here, we focus on the oleic acid ligand capping layer of the NPs and its oxidation and partial volatilization by reactive oxidizers. The $NaYF_4$ cores act both as a support for the oleic acid ligands and also as an X-ray photon absorber and hence generator of slow secondary electrons, which are known to generate highly reactive oxidation agents, such as OH radicals and $O_3$, in the presence of $O_2$ and residual water vapor. These species then drive the degradation of the oleic acid ligands through oxidation reactions, leading to a shrinkage of the ligand layer through oxidative volatilization. At the same time, the chemical nature of the ligand layer changes due to the formation of new functional groups by oxygenation of the oleic acid molecules.

Ambient pressure X-ray photoelectron spectroscopy (APXPS) was used in past investigations to monitor the reaction of carbonaceous species (specifically coronene) with highly-reactive oxidizing trace gases, such as OH radicals and $O_3$.[11] The chemical and surface sensitivity of XPS allowed to distinguish different functional groups in C 1s spectra and also to determine the total carbon loss through volatilization of coronene and its oxidation products. One complication in this previous study, however, was that the morphology of the reacted film could not be monitored directly, which introduces uncertainties in the analysis of the APXPS data. In the present work we address this issue by combining APXPS measurements with grazing incidence X-ray scattering (GIXS) *in situ*, which provides information about morphological changes to the NPs and their oleic acid ligand layer and is thus complementary to APXPS.[12]

The results of our APXPS and GIXS study shows that oxidation of the oleic acid layer leads to the formation of alcohol, carbonyl and acid groups, and that the overall volatilization of oleic acid is the dominating process. GIXS data indicate that the reaction preferentially takes place on the side of the NPs facing the gas phase, ultimately resulting in a chemically and morphologically asymmetric ligand layer, where the oxidized part shows mainly $CH_x$ and acid group carbons. We believe that the present investigation is a model for in-depth studies of a wide range of reactions which affect both the chemical nature as well as the morphology of ligand-capped nanomaterials and other carbonaceous nanoscale systems, using complementary, surface sensitive spectroscopic (APXPS) and scattering (GIXS) methods.

## 2. RESULTS AND DISCUSSION

The oxidation of the oleic acid layer surrounding the 9 nm $NaYF_4$ nanoparticles was monitored using APXPS as a function of $O_2$ background pressure and time. For each measurement a freshly-prepared sample was used. The as-prepared samples were first characterized in vacuum by XPS and GIXS. Afterwards, $O_2$ was admitted to the sample compartment and the reaction was monitored over time using XPS, with a focus on the C 1s spectra which report on the formation of oxygenated products and the overall volatilization of the carbonaceous layer over time. While an external ozone generator could be used to facilitate the oxidation reaction, it turned out that photodissociation of $O_2$ (and residual $H_2O$) at sub-mbar pressure, as well as electron-impact excitation and dissociation of $O_2$ (and likely residual $H_2O$) produced a sufficient amount of reactive oxygenated species for the volatilization of the oleic acid NP layer.

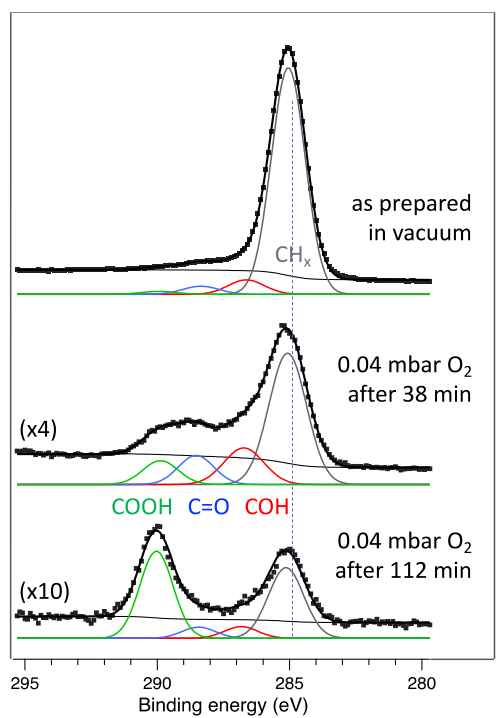

**Figure 1.** C 1s spectra of $NaYF_4$ NPs surrounded by an oleic acid layer and deposited on a SiOx/Si wafer. The top spectrum was taken in vacuum before oxidation. It is dominated by the $CH_x$ peak of oleic acid, with small traces of oxygenated species already present, which can possibly also arise from adventitious carbon on the substrate. The middle and bottom spectra were recorded during the X-ray beam-assisted oxidation in 0.04 mbar $O_2$. The growth of the oxidation products (COH, C=O, COOH) can be observed alongside the reduction of the $CH_x$ peak. The BE axis was corrected so that the $CH_x$ peak of the as-prepared sample is at 285 eV, the literature value. Please note the different intensity scaling of the spectra.

## 2.1 Ambient pressure X-ray photoelectron spectroscopy

Figure 1 shows C 1s spectra taken before (top), during (middle) and at the end (bottom) of the oxidation reaction, here for 9 nm $NaYF_4$ NPs in 0.04 mbar of $O_2$. The spectrum of the as-prepared sample is dominated by the $CH_x$ peak (at 285 eV), as expected from the chemistry of oleic acid, i.e., $C_{17}H_{33}$-COOH. Some amount of oxidized carbon is also present on the sample, which could be due to adventitious carbon, and to a smaller part due to the acid groups of oleic acid, which have a BE of ~290 eV.

Once $O_2$ is introduced to the chamber, the oxidation of oleic acid (and the adventitious carbon) proceeds readily, as can be seen in the spectra taken after 38 min and

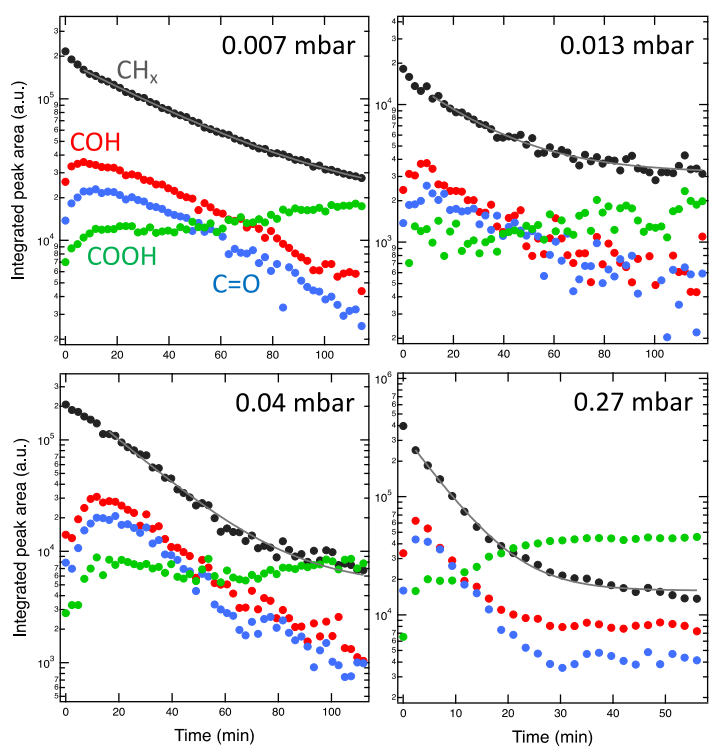

**Figure 2.** Integrated peak area for the four component peaks in the C 1s spectrum as shown in Figure 1 during oxidation at four different $O_2$ pressures. In each case a continuous decrease of the $CH_x$ species is observed, while the oxygenated products first grow in intensity and then decrease as well. The solid lines are exponential fits of the decay in the $CH_x$ signal. Please note that the time and peak area axis do not have the same range for the four different plots.

112 min in the presence of $O_2$. Three product peaks can be clearly distinguished, with binding energies consistent with those for alcohol (~286.5 eV), carbonyl (~288 eV) and carboxylic acid (~290 eV) groups.[13] The temporal evolution of these species alongside the reduction of the amount of $CH_x$ is shown in Fig. 2 for experiments at four different $O_2$ pressures between 0.007 mbar and 0.27 mbar. It is apparent that there is a generational evolution of the different products in the order of COH – C=O – COOH, as previously observed for the case of the oxidation of coronene.[11] As the oxidation advances, the COOH species become the most abundant ones, together with some residual $CH_x$. The ratio of $COOH/CH_x$ at the end of the oxidation depends on the $O_2$ pressure, as does the rate of the oxidation, which increases with $O_2$ pressure.

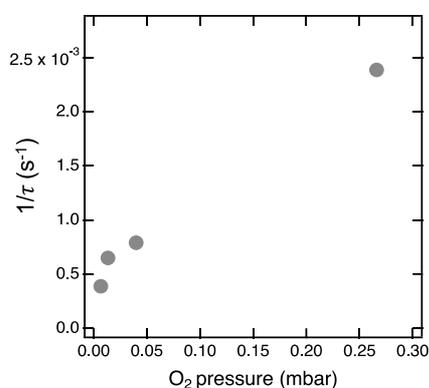

**Figure 3.** Inverse time constant of the decay of the $CH_x$ peak as a function of $O_2$ background pressure.

To quantify the dependence of the reaction rate on the pressure we have fitted an exponential dependence to the temporal evolution of the $CH_x$ signal, shown as solid lines in Fig. 2. The inverse time constants determined from these fits are plotted in Fig. 3 as a function of the $O_2$ pressure. There is a linear dependence between these two quantities, which implies a first order dependence of the reaction rate on the concentration of reactive oxygen species. This is reasonable considering that the availability of reactive species directly depends on the number of $O_2$ molecules in the gas phase, no matter if the process of their creation is electron impact ionization due to secondary, Auger and photoelectrons from the sample and gas, or direct photoionization of the gas phase by the incident X-rays. All other conditions, such as temperature, incident photon flux and energy, and sample composition (which is important for the number and energy distributions of electrons) are similar across these experiments. Plots of the total carbon content as well as the total carbon and oxygen content in the ligand layer as a function of reaction time (see Figure S9 in the SI) show that volatilization is the dominating process and growth of the ligand layer through functionalization in the early stages of oxidation seem to be negligible, unlike in the case of coronene[11]. The

C/O ratio vs time approaches unity (Figure S10 in the SI) in all cases, which is in line with the formation of oxidation products with an average stoichiometry akin to that in acidic acid.

## 2.2 Grazing incidence X-ray scattering

We now turn our attention to the investigation of the morphology of the NPs and the oleic acid ligand layer. *Ex-situ* AFM measurements on as-prepared samples (Figure 4) reveal a coverage of approximately one monolayer of the NPs in a closely packed configuration. Occasionally some NPs are observed in the second layer, confirmed also by the *in situ* GIXS measurements shown later. The mean particle height and the inter-particle distances determined by AFM are 10 nm and 13 nm, respectively, which is a reasonable value for 9 nm $NaYF_4$ NPs covered by oleic acid ligands.

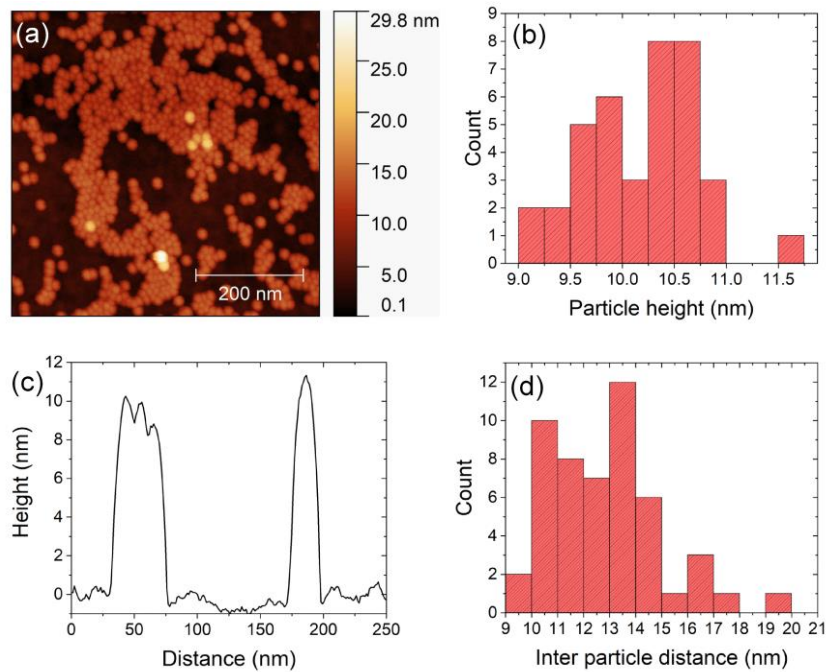

**Figure 4.** (a) Topography of the as-prepared sample used in the reaction a 0.007 mbar $O_2$, characterized by AFM. The NPs are located in two layers, where the bottom layer has a coverage of ~45%. The 2nd layer (height between 10 and 20 nm) comprises mostly isolated single NPs. (b) The average height of the NPs in the bottom layer, as determined by AFM, is 10.2 ± 0.5 nm. (c) Topographic line scan showing the profile of a single NP (at ~170 nm) and of clustered NPs (at ~50 nm), where the NP shape is convoluted with that of the AFM tip. (d) Distribution of the in-plane separation of NPs, showing an average particle spacing of 12.9 ± 2.1 nm.

*In situ* GIXS patterns measured before and after oxidation in the presence of 0.007 mbar $O_2$ are shown in Figure 5 (a) and (b). The changes in the pattern after oxidation are pronounced and manifest themselves in an overall broadening of the vertical rods, which is mostly due to changes in the structure factor. Changes are also observed for the Yoneda line, which is a horizontal high-intensity line observed at the critical angle, indicated by arrows in Fig. 5(a).

An analytical fit of line-cuts along the Yoneda line utilizing a core-shell form factor model,[14] a structure factor for a face centered cubic (fcc) pattern [15,16] and a fractal from factor[17] reveals a decrease of the overall NP size from 13 nm to 11 nm upon oxidation (see Figures 5(c) and (d)). Please note that the analysis at $q_r$ values below $4\times10^{-2}$ nm$^{-1}$ is not possible due to the presence of the zero order beam stop, and thus data points for these $q$ values are not included in the fit. The model used to represent the sample is based on closely-packed spherical capped nanoparticles with a nominal diameter of the $NaYF_4$ core of 9.6 nm. The decrease of the ligand layer thickness is a direct consequence of the volatilization of oleic acid and its reaction products during the oxidation, as observed by APXPS (see Figs. 1 and 2). The overall broadening of the features in the GIXS map of the reacted sample points to an increase in the polydispersity, both in terms of size and shape of the NPs and the spacing between them. More details on the fit methodology are available in Section 1 (Analytical model description) of the Supplementary Information.

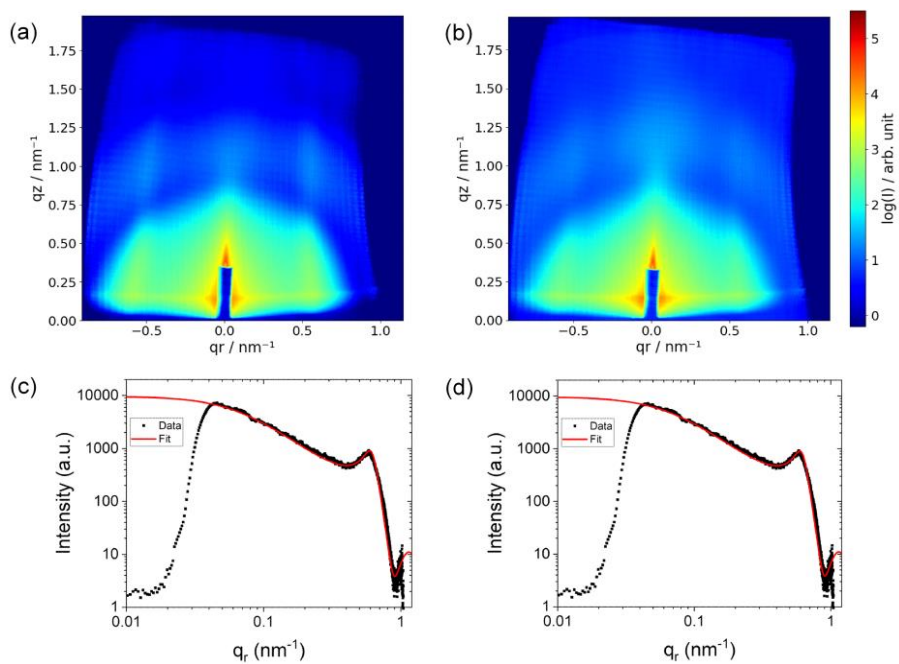

**Figure 5.** **(a)** and **(b):** X-ray scattering pattern of NaYF$_4$/oleic acid ligand NPs measured with an incident photon energy of 1240 eV before and after oxidizing treatment in 0.007 mbar, respectively. The nominal diameter of the NPs is 9.6 nm. **(c)** and **(d):** Analytical fit of a line-cut through the scattering data along the Yoneda line (indicated in A by two arrows) of patterns shown in (a) and (b). The scattering data are modeled assuming closely packed spheres with diameters of 13 nm and 11 nm, respectively. The q-range below ~0.5 nm$^{-1}$ is dominated by the NPs' form factor, while the range above 0.5 nm$^{-1}$ is due to the structure factor. The reduction of the NPs size from 13 nm to 11 nm models the reduction of the oleic acid ligand layer thickness due to the oxidation reaction.

We have performed the fit analysis shown in Fig. 5 for the case of the measurements at 0.007 mbar also for the other pressures. The reduction in the thickness of the organic ligand layer and the interparticle distance on the O$_2$ pressure is shown in Figure 6. The data in the left panel imply that the reduction in organic layer thickness approaches a limiting value at the highest O$_2$ pressure of 0.27 mbar. This is in agreement with the APXPS data in Fig. 2, which show for the 0.27 mbar oxidation a steady state of the C 1s components already halfway through the experiment, after about 30 min.

The analysis of the interparticle distances after the reaction based on GIXS measurements, shown in Fig. 6(b), indicate that the NPs appear to be firmly bonded to the SiO$_x$ substrate still maintain their oleic acid ligand layer in the horizontal plane, since the original in-plane interparticle distance of ~13.5 nm is still observed after the reaction. This distance is in good agreement with the initial ex situ

AFM measurements (Figure 4). This observation also implies that most of the changes to the oleic acid ligand layer is taking place in the out-of-plane direction, as will be discussed next.

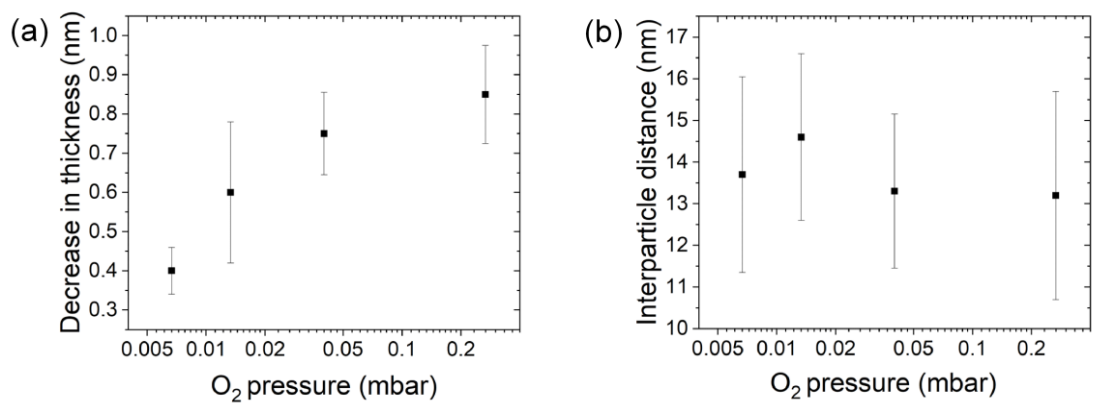

**Figure 6.** (a) Average reduction in the thickness of the carbonaceous layer around the $NaYF_4$ core. The error bars are based on the polydispersity. (b) Interparticle distance (solid squares) and particle displacement (error bars) in the particle fcc lattice during $O_2$ treatment as a function of pressure. The data are extracted from in-plane 1D fits of the Yoneda line, as shown for the 0.007 mbar $O_2$ measurement in Figure 5D. The equivalent fits for the other $O_2$ pressures are shown in the SI in the section "Analytical model description" and in Figures S1 and S2.

We now turn our attention to the analysis of the measured X-ray scattering data by comparing them with theoretically predicted scattering patterns based on the simulation of structural models for the NP size, shape and distribution. Figure 7 shows the analysis on the example of the sample oxidized in 0.007 mbar $O_2$. In Fig. 7(a), the experimental data are displayed, while Fig. 7(b) shows calculated data utilizing the BornAgain simulation package.[18,19,20] The simulated GIXS map in Fig. 7(b) based on our model (discussed in the next paragraph) shows all the characteristic features of the experimental data, namely a strong enhancement of the in-plane signal along the Yoneda line, vertical Bragg rods originating from the in-plane structure factor, and a diffuse ring arising from the 3D form factor of the ligand-covered NPs.

The quantitative comparison of theoretical and experimental data is based on the line cuts along the horizontal Yoneda line and the vertical scattering plane. This quantitative comparison is shown in Figs. 7(c) and (d). The agreement between simulated and measured data is very good, especially for larger $q_r$ and $q_z$ values, which mainly reflect the properties of the individual NPs and their distances from each other. However, due to restrictions in the input geometry of the state of the art GIXS simulation packages,[20,21,22] the reconstructed data can reproduce experimental data only partially, as fcc packing distribution, variations in particle-substrate distance, and particle clustering cannot be simulated alongside particle size distributions.[23,24] For example, deviation for lower $q_z$ (out-of-plane direction, Figure 7c) is due to the reflectivity contribution and the roughness of the substrate, not included in the numerical simulations. More details on sensitivity of fit between simulated and experimental data on specific model parameters are shown in the Section 2 of Supplementary Information (Figs. S3-S8).

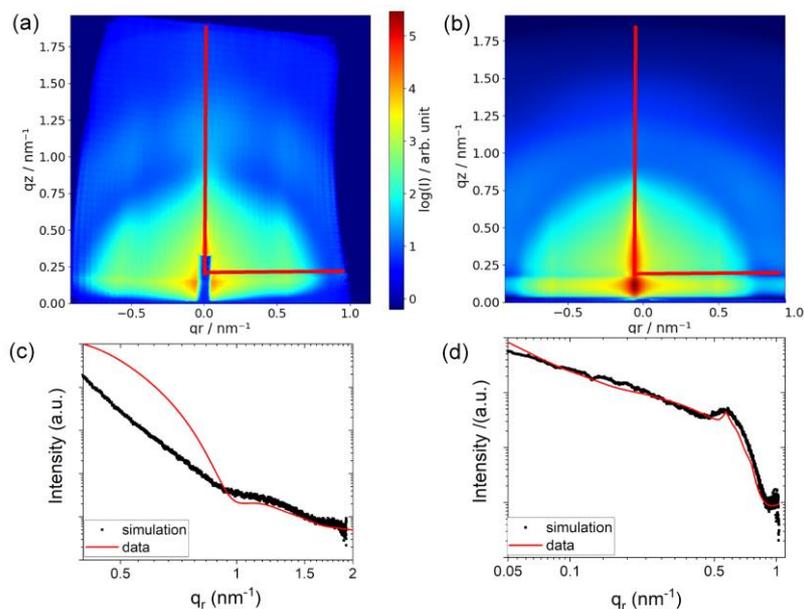

**Figure 7.** (a) Measured and (b) simulated X-ray scattering pattern of the sample after oxidation in 0.007 mbar $O_2$. (c) Horizontal line cuts along the Yoneda line and (d) vertical line cuts along the scattering plane of the experimental and simulated data showing quantitative agreement in the position of the minima and maxima of the line cuts and their slope. The position of the line cuts are indicated in (a) and (b). The black lines in (c) and (d) are the experimental data, the red lines the results of the simulation.

The details of the proposed model are discussed next. Initially, the substrate is covered with a 4 nm thick (possibly non-continuous) layer of a carbonaceous layer with a mean/effective density of 1/4 of the density of the nominal oleic acid value. This layer is formed by the oleic acid excess and the NPs closest to the substrate are submerged in this film (SI Figure S3C). Based on the simulation results, the mass density of the oleic acid in the ligand layer is similar to the density of pure oleic acid[25], while the density of the $NaYF_4$ nanoparticles was found to be 20% lower than in literature reports.[26]

Figure 8 shows the model used for 0.007 mBar $O_2$ treated sample after oxidation. The polydispersity and the size of nanoparticles determined from X-ray scattering (9.4 ± 0.6 nm) is in excellent agreement with NP dimensions determined in scanning transmission electron microscopy (not shown) measurements (9.4 ± 0.4 nm). More details on how the individual layers in the Figure 8 model were defined are in the Supplementary information - Section 3.

During oxidation the thickness of the carbonaceous layer covering the $NaYF_4$ core decreases from 1.3 nm (as-prepared oleic acid layer) to 0.8 nm (oxidation products). The majority of carbonaceous material is removed at the top of the particles, while the ligand layer at the bottom of the NPs (in contact with the substrate) shows little thinning (see Fig. 8a). During the oxidation process, the adventitious carbon layer initially covering the substrate is also fully removed, according to the results of the comparison of simulated and measured data.

Based on the analysis of the in-plane lattice spacing between NPs (as shown in Fig. 8b), there are several populations present both before and after oxidation. In the as-prepared sample, the majority of the NP species exhibits a mean in-plane distance of 12.9 nm and a lattice displacement of 1.25 nm; a minority population of the as-prepared NPs exhibits a broad range of spacing ranging from 10 nm to 30 nm with large relative displacements (more than 10%), creating a pseudo-randomly ordered population. After oxidation, the in-plane lattice spacing of the majority NP population decreases to 12.5 nm and the particle polydispersity increases from 0.5 to 0.7 (Figure 8c). For both as-prepared and oxidized sample some NP pairs are observed that share a carbonaceous layer. This is, however, only a small fraction (~1%) of the total NP population.

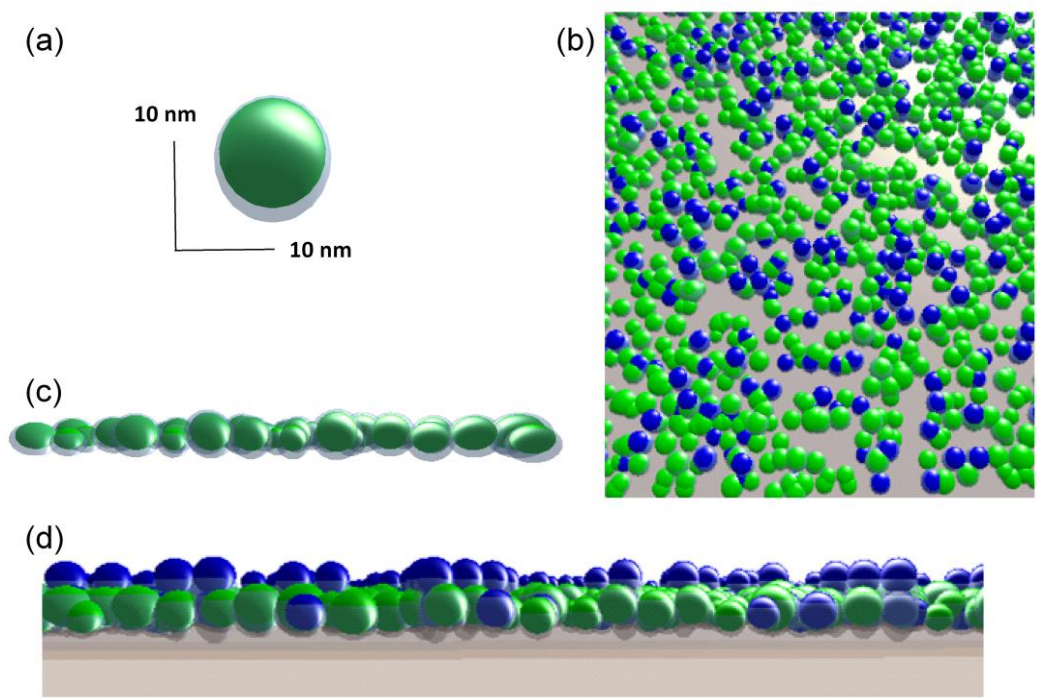

**Figure 8.** (a) Visualization of a single NP with capping layer, after oxidation. The asymmetrical shape of the capping layer is a consequence of exposure to $O_2$, with limited reaction rates in plane and at the contact to the substrate. (b) 3D model used to simulate the scattering pattern. Particles in the 1st layer are closely packed in a quasi-regular hexagonal pattern with 45% coverage. The 2nd layer is partially filled with quasi-regular in-plane stacking. All $NaYF_4$ particles are fully covered by an oleic acid ligand layer with a thickness of 0.6 nm thickness. Due to clustering and other changes of morphology, the NPs have a distribution of diameters, simulated by a Gaussian with ±1 nm FWHM. (c) Graphical depiction of the particle size distribution utilized for the simulation. (d) Side View of the simulation geometry. Particles are lifted off the substrate as they are suspended in the excess capping agent layer. The height of the different fcc stacking is varied between 2.5 nm (blue) and 0.6 nm (green).

From the analysis of the out-of-plane line-cuts it is deduced that there is a relatively large range of substrate-NPs distances. This is caused by the formation of a second layer of NPs on top of the monolayer, as well as by the asymmetrical position of some of the top-most NPs within their ligand shell after oxygen exposure. The preferred configuration of the NPs in the second layer is at the bridge site between pairs of particles in the bottom layer. The bottom layer of particles varies in their positions over the substrate between 0 nm and 6 nm (Figure 8d). This distance variation is substantially reduced after $O_2$ exposure to 3 nm, due to the volatilization of the excess oleic acid layer during oxidation.

Despite the simplifications used in the model and the existence of other structures that yield similar scattering patterns, the simulated particle sizes, polydispersity, thickness of capping agent layer, average in plane particle distance and the existence of two layers are dependent on highly indicative features like the position and width of the powder ring, the $q$ position of the structure factor rods and the slope around the first minimum of the form factor. Therefore, and due to the good agreement between simulated data and the analytical approach, we are confident that the parameters mentioned above are correctly reflected in the simulation of the scattering data.

## 3. CONCLUSIONS

A model core-shell nanoparticle composed of inorganic $NaYF_4$ core and oleic acid ligand layer was studied *in situ* by complementary APXPS and GIXS, revealing the chemical and morphological transformation of the ligand layer due to oxidation. From APXPS data it was determined that the reaction rate depends linearly on the $O_2$ partial pressure, suggesting a first-order transformation of oleic acid into several reaction intermediates, including alcohol, carbonyl and carboxylic acid groups, the latter ones representing the majority species at the end of the reaction. Most of the ligand film is removed during the oxidation process due to volatilization. GIXS data indicate that during the reaction the NPs remain firmly attached to the Si substrate, while losing appreciable thickness of the oleic acid layer at the side of the NPs facing the gas atmosphere. It was also found that oxidation leads to a restructuring of the NPs, lowering their mean distance. The effect was observed both for in-plane and out-of-plane distances. In addition, increased clustering of NPs due to oxidation was observed.

The approach presented in this work is applicable to a wide variety of nanoparticles and other monodisperse systems, such as emulsions. The combination of the APXPS and GIXS, which provides information on morphological and chemical changes, holds great promise for future multi-modal studies in the fields of environmental science, heterogeneous catalysis, photo- and electrochemistry, and beyond.

## 4. METHODS

*Nanoparticle synthesis and sample preparation*

$NaYF_4$: 8% $Tm^{3+}$ upconverting nanoparticles (UCNPs) with a mean diameter size of 9.4 nm were synthesized based on a previously reported method.[27] For a typical synthesis, $YCl_3$ (0.92 mmol, 180 mg) and $TmCl_3$ (0.08 mmol, 22 mg) were added into a 50-ml three-neck flask, followed by an addition of 6 ml OA and 14 ml ODE. The solution was stirred under vacuum and heated to 100 °C for 1 h. During this time, the solution became clear. After that, the flask was then subjected to three pump–purge cycles, each consisting of refilling with $N_2$ and immediately pumping under vacuum to remove water and oxygen. Afterwards, sodium oleate (2.5 mmol, 762 mg) and $NH_4F$ (4 mmol, 148 mg) were added to the flask under $N_2$ flow. Subsequently, the resealed flask was resealed and placed under vacuum for 15 min at 100 °C, followed by three pump–purge cycles. At the end of the treatment, the flask was quickly heated from 100 °C to 320 °C (the approximate ramp rate was 25 °C min$^{-1}$). The temperature was held at 320 °C for 40 min, after which the flask was rapidly cooled to room temperature using a stream of compressed air.

To isolate the nanoparticles, ethanol was added to the solution in a 1:1 volume ratio, and the precipitated nanoparticles were isolated by centrifugation (5 min at 4,000 rpm). The pellet was suspended in hexanes and centrifuged to remove large and aggregated particles. The nanoparticles remaining in the supernatant were washed two additional times by adding ethanol, isolating by centrifugation and dissolving the pellet in hexanes. The nanoparticles were stored in hexanes.

The NPs were deposited onto Si wafers by immersing a Si wafer piece with dimensions of ~5 mm x 20 mm x 0.2 mm face down into the solution, which was kept in a 5 mL screw cap bottle lying on its side. In this manner the Si surface did not touch the walls of the tube (only at the outer edges) and the NPs assembled from solution onto the Si surface through adhesive interactions and not gravity. The Si wafers were left immersed in the NP/hexane solutions for a minimum of 10 minutes before being removed, dried in air and mounted inside the APXPS instrument without further treatment.

*APXPS and GIXS measurements*

The APXPS and GIXS measurements were performed using the APPEXS (Ambient Pressure PhotoElectron spectroscopy and X-ray Scattering) setup at the APXPS-2 port of beamline 11.0.2 of the Advanced Light Source at Lawrence Berkeley National Laboratory in Berkeley, CA.[12] The incident photon energy was 1000 eV for the APXPS and 1240 eV for the GIXS maps, with an exit slit size of 60 $\mu$m × 250 $\mu$m. The resulting photon flux at these energies is estimated to be 3×10$^{11}$ photons/sec and 6×10$^{10}$ photons/sec, respectively. The combined beamline and electron analyzer resolution in the experiments was better than 0.8 eV. The X-rays were incident on the sample under a grazing angle of ~2 deg in the proximity of the critical angle of the Si substrate, with the electron detection direction at 15 deg from the sample normal. X-ray scattering data were collected using a 2D CCD detector (Andor iKon-L) mounted on a two-axis camera rotating manipulator covering a solid angle of ±12 deg in-plane and +24 deg out-of-plane (above the horizon). This corresponds to covering a scattering vector $q$-range of ±1.5 nm$^{-1}$ and 3nm$^{-1}$, respectively. The composite scattering image was reconstructed from 1476 individual images with resulting $q$-resolution of 5×10$^{-4}$ nm$^{-1}$. The X-ray detector was separated from the reaction chamber by a large-area ultra-thin (150 nm) $Si_3N_4$ window matching the area of the detector (27×27 mm$^2$), assuring that the pressure near the CCD chip is kept below 10$^{-6}$ Torr and the X-ray transmission is sufficiently high even in soft X-ray regime.

The emphasis of the APXPS measurements was on the C 1s, Y 3p, O 1s and Si 2p core levels. C 1s and Y 3p (which have very similar binding energy) report on the chemical state of the NP shell and core, respectively, while Si 2p reveals changes to the coverage of the Si substrate by the NPs and possible oxidation of the Si wafer. The O 1s signal stems from both the thin oxide layer on the substrate and the reaction products of the oxidation of oleic acid and any residual carbon contamination on the Si wafer. The XPS data were deconvoluted using a commercial software package (KolXPD). Pure Gaussians were sufficient to fit all peaks.

## ASSOCIATED CONTENT

### Data Availability Statement

All relevant data and analysis scripts used in this study are available from the corresponding authors upon reasonable request.

### Supporting Information

The Supporting Information is available free of charge. It contains details on the analysis of the GIXS data.

## AUTHOR CONTRIBUTIONS

S.N., M.J., K.R.W and H.B. designed the research and experiments. X.Q. and E.C. synthesized the NP samples. S.N., M.J. and H.B. carried out the experiments at the ALS. M.J. and S.N. analyzed the GIXS and H.B. the XPS data. S.N., M.J., and H.B. wrote the manuscript with input from all authors.

### Notes

The authors declare no competing financial interest.

## ACKNOWLEDGEMENTS

This work was supported by the Condensed Phase and Interfacial Molecular Science Program (CPIMS), in the Chemical Sciences Geosciences and Biosciences Division of the Office of Basic Energy Sciences of the U.S. Department of Energy under Contract No. DE-AC02-05CH11231. M.J and M.S were supported by the Catalysis program FWP CH030201. The Advanced Light Source is supported by the Director, Office of Science, Office of Basic Energy Sciences of the U.S. Department of Energy at LBNL under Contract No. DE-AC02-05CH11231. Work at the Molecular Foundry was supported by the Office of Science, Office of Basic Energy Sciences, of the U.S. Department of Energy under Contract No. DE-AC02-05CH11231. Authors thank Virginia Altoe and Paul Ashby for their help with microscopy measurements.

# REFERENCES

[1] A. Heuer-Jungemann, N. Feliu, I. Bakaimi, M. Hamaly, A. Alkilany, I. Chakraborty, A. Masood, M.F. Casula, A. Kostopoulou, E. Oh, K. Susumu, M.H. Stewart, I.L. Medintz, E. Stratakis, W.J. Parak, and A.G. Kanaras, The Role of Ligands in the Chemical Synthesis and Applications of Inorganic Nanoparticles, *Chem. Rev.* 2019, 119, 8, 4819–4880

[2] J.H. Kroll, N.M. Donahue, J.L. Jimenez, S.H. Kessler, M.R. Canagaratna, K.R. Wilson, K.E. Altieri, L.R. Mazzoleni, A.S. Wozniak, H. Bluhm, E.R. Mysak, J.D. Smith, C.E. Kolb, D.R. Worsnop, Carbon oxidation state as a metric for describing the chemistry of atmospheric organic aerosol, *Nature Chemistry* 3, 133-138 (2011).

[3] X. Xia, E. Sivonxay, B.A. Helms, S.M. Bau, E.M. Chan, Accelerating the Design of Multishell Upconverting Nanoparticles through Bayesian Optimization, *Nano Lett.* 23, 11129-11136 (2023).

[4] A.D. Ostrowski, E.M. Chan, D.J. Gargas, E.M. Katz, G. Han, P.J. Schuck, D.J. Milliron, B.E. Cohen, Controlled Synthesis and Single-Particle Imaging of Bright, Sub-10 nm Lanthanide-Doped Upconverting Nanocrystals, *ACS Nano* 6, 2686-2692 (2012).

[5] C. Lee, E.Z. Xu, Y. Liu, A. Teitelboim, K. Yao, A. Fernandez-Bravo, A.M. Kotulska, S.H. Nam, Y.D. Suh, A. Bednarkiewicz, B.E. Cohen, E.M. Chan, P.J. Schuck, Giant nonlinear optical responses from photon-avalanching nanoparticles, *Nature* 589, 230–235 (2021).

[6] C. Lee, E.Z. Xu, K.W.C. Kwock, A. Teitelboim, Y. Liu, H.S. Park, B. Ursprung, M.E. Ziffer, Y. Karube, N. Fardian-Melamed, C.C.S. Pedroso, J. Kim, S.D. Pritzl, S.H. Nam, T. Lohmueller, J.S. Owen, P. Ercius, Y.D. Suh, B.E. Cohen, E.M. Chan, P.J. Schuck, Indefinite and bidirectional near-infrared nanocrystal photoswitching, *Nature* 618, 951–958 (2023).

[7] A. Skripka, M. Lee, X. Qi, J.-A. Pan, H. Yang, C. Lee, P.J. Schuck, B.E. Cohen, D. Jaque, E.M. Chan, A Generalized Approach to Photon Avalanche Upconversion in Luminescent Nanocrystals, *Nano Lett.* 23, 7100–7106 (2023).

[8] Y. Liu, A. Teitelboim, A. Fernandez-Bravo, K. Yao, M.V.P. Altoe, S. Aloni, C. Zhang, B.E. Cohen, P.J. Schuck, E.M. Chan, Controlled Assembly of Upconverting Nanoparticles for Low-Threshold Microlasers and Their Imaging in Scattering Media, *ACS Nano* 14, 1508–1519 (2020).

[9] X. Ou, X. Qin, B. Huang, J. Zan, Q. Wu, Z. Hong, L. Xie, H. Bian, Z. Yi, X. Chen, Y. Wu, X. Song, J. Li, Q. Chen, H. Yang, X. Liu, High-resolution X-ray luminescence extension imaging, *Nature* 590, 410–415 (2021).

[10] M.B. Prigozhin, P.C. Maurer, A.M. Courtis, N. Liu, M.D. Wisser, C. Siefe, B. Tian, E.M. Chan, G. Song, S. Fischer, S. Aloni, D.F. Ogletree, E.S. Barnard, L.-M. Joubert, J. Rao, A.P. Alivisatos, R.M. Macfarlane, B.E. Cohen, Y. Cui, J.A. Dionne, S. Chu, Bright sub-20-nm cathodoluminescent nanoprobes for electron microscopy, *Nature Nanotechnology* 14, 420–425 (2019).

[11] E.R. Mysak, J.D. Smith, J.T. Newberg, P.D. Ashby, K.R. Wilson, H. Bluhm, Competitive Reaction Pathways for Functionalization and Volatilization in the Heterogeneous Oxidation of Coronene Thin Films by Hydroxyl Radicals and Ozone, *Phys. Chem. Chem. Phys.* 13, 7554 (2011).

[12] H. Kersell, P. Chen, H. Martins, Q. Lu, F. Brausse, B.-H. Liu, M. Blum, S. Roy, B. Rude, A.D. Kilcoyne, H. Bluhm, S. Nemšák, Simultaneous ambient pressure X-ray photoelectron spectroscopy and grazing incidence X-ray scattering in gas environments, Rev. Sci. Instrum. 92, 044102 (2021).

[13] G. Beamson and D. Briggs, *High resolution XPS of organic polymers: The Scienta ESCA300 database*. 1992, Chichester [England], New York: Wiley.

[14] A Guinier and G Fournet, Small-Angle Scattering of X-Rays, John Wiley and Sons, New York, (1955).

[15] H. Matsuoka, H. Tanaka, T. Hashimoto, and N. Ise, Elastic scattering from cubic lattice systems with paracrystalline distortion, *Physical Review B*, 36 (1987) 1754-1765

[16] H. Matsuoka, H. Tanaka, N. Iizuka, T. Hashimoto, and N. Ise, Elastic scattering from cubic lattice systems with paracrystalline distortion. II, *Physical Review B*, 41 (1990) 3854-3856

[17] J. Teixeira, Small-angle scattering by fractal systems, J. Appl. Cryst., 21 (1988) 781-785

[18] G. Pospelov, W. Van Herck, J. Burle, J. M. Carmona Loaiza, C. Durniak, J.M. Fisher, M. Ganeva, D. Yurova and J. Wuttkea. BornAgain: software for simulating and fitting grazing-incidence small-angle scattering, J. Appl. Cryst. 53, 262–276 (2020)

[19] A. Nejati, M. Svechnikov and J. Wuttke, BornAgain, software for GISAS and reflectometry: Releases 1.17 to 20, EPJ Web Conf. 286, 06004 (2023)

[20] https://www.bornagainproject.org

[21] S. T. Chourou, A. Sarje, X. S. Li, E. R. Chan and A. Hexemer, HipGISAXS: a high-performance computing code for simulating grazing-incidence X-ray scattering data, *J. Appl. Cryst.* (2013). 46, 1781-1795.

[22] R. Lazzari, IsGISAXS: a program for grazing-incidence small-angle X-ray scattering analysis of supported islands, *J. Appl. Cryst.* 35 (2002) 406-421.

[23] Wuttke, J., Cottrell, S., Gonzalez, M.A., Kaestner, A., Markvardsen, A., Rod, T.H., Rozyczko, P., Vardanyan, G., Guidelines for collaborative development of sustainable data treatment software *J. Neutron Res.* 24, 33-72 (2022)

[24] J. Wuttke, Numerically stable form factor of any polygon and Polyhedron, *J. Appl. Cryst.* 54, 580-587 (2021).

[25] A. Thomas, *Fats and Fatty Oils*, in: Ullmann's Encyclopedia of Industrial Chemistry, Wiley-VCH (2011).

[26] L. E. Mackenzie, J. A. Goode, A. Vakurov, P. P. Nampi, S. Saha, G. Jose, and P. A. Millner, The theoretical molecular weight of NaYF$_4$:RE upconversion nanoparticles, *Scientific Reports* 8, 1106 (2018).

[27] Ostrowski, A. D. et al. Controlled synthesis and single-particle imaging of bright, sub-10 nm lanthanide-doped upconverting nanocrystals. ACS Nano 6, 2686–2692 (2012).

# Direct Observation of Morphological and Chemical Changes During the Oxidation of Model Inorganic Ligand-Capped Particles

*Supplementary Information*

Maximilian Jaugstetter[1,*], Xiao Qi[2], Emory Chan[2], Miquel Salmeron[1], Kevin R. Wilson[3], Slavomír Nemšák[4,5,*], Hendrik Bluhm[6,*]

[1] *Materials Sciences Division, Lawrence Berkeley National Laboratory, Berkeley, CA 94720, USA*

[2] *Molecular Foundry, Lawrence Berkeley National Laboratory, Berkeley, CA 94720, USA*

[3] *Chemical Sciences Division, Lawrence Berkeley National Laboratory, Berkeley, CA 94720, USA*

[4] *Advanced Light Source, Lawrence Berkeley National Laboratory, Berkeley, CA 94720, USA*

[5] *Department of Physics and Astronomy, University of California, Davis, CA 95616, USA*

[6] *Fritz Haber Institute of the Max Planck Society, D-14195 Berlin, Germany*

## *Section 1 – GIXS analytical model description*.

In-plane cuts were performed at $q_z = 0.2$ nm$^{-1}$, which corresponds to the critical angle of 2° of the NaYF$_4$ nanoparticles to maximize the contribution of the nanoparticle to the scattering curve. The critical angle of NaYF$_4$ was calculated using Henke tables[1] at a photon energy of 1240 eV using a chemical formula of NaYF$_4$ and density of 5.578 gcm$^{-3}$ as calculated from the molecular weight data given by Mackenzie et al.[2]

The model used to fit in-plane line cuts (along Yoneda line) consists of a core/shell sphere form factor.[3] Here, the NaYF$_4$ nanoparticles function as the sphere's core and the oleic acid ligand layer as spheres shell. Size distribution is added via a Gaussian distribution model with set average and standard deviation defined by the PD value. A fractal structure factor[4] consisting of the spherical from factor of the solid nanoparticles as single unit and a total equivalent radius of 24 nm with a dimensionality of 2.2 to account for formed pseudo-crystalline assemblies of the deposited nanoparticles. Lastly, 3 FCC structure factors with different spacing[5] were used to account for the different populations in the formed nanoparticle islands (as shown in figure 4 in the main text) on the substrate. Different spacing is given for top layer with a weight of around 10 %, for the interference between particles at the edges of the formed island with another 10 % contribution.

The main structure factor can be observed as shoulders in all measured scattering patterns and appear as peaks in the individual line cuts and is defined by the inter-particle distance within the individual islands. This structure factor contributes 80 % to the total structure factor and shows slightly bigger inter-particle distance than particle size with varying spacing between the samples (see Figure 6 in the main text). Individual line-cuts together with fits are shown in the main text in Figure 5 and Figures S1, S2 and S3.

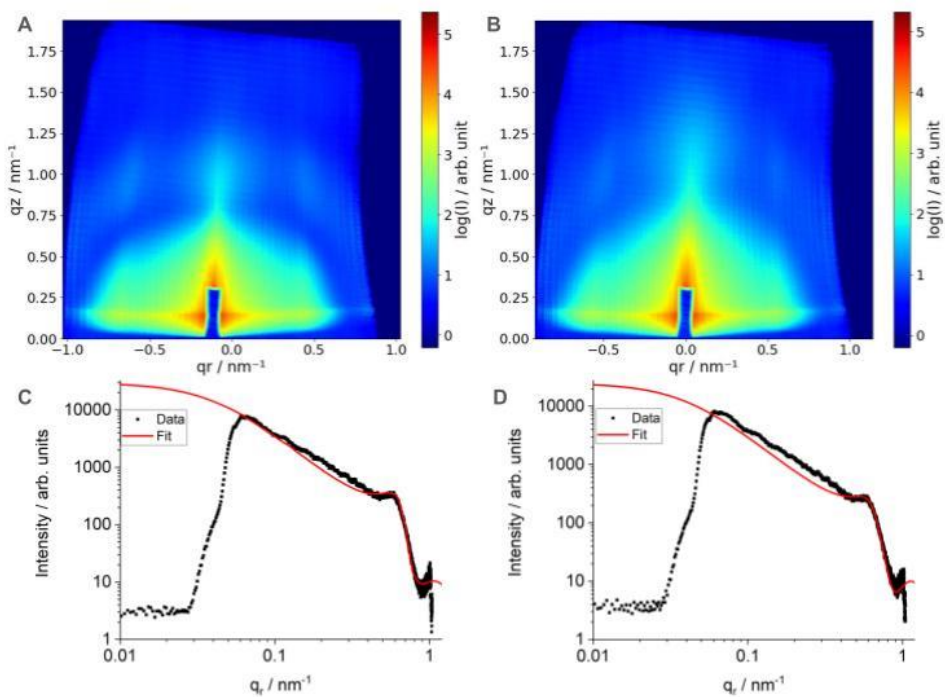

**Figure S1:** Scattering pattern (**A, B**) and fitted in-plane line (**C, D**) cuts for NaYF$_4$ particles capped with oleic acid before (left) and during treatment with 0.013 mBar O$_2$ (right).

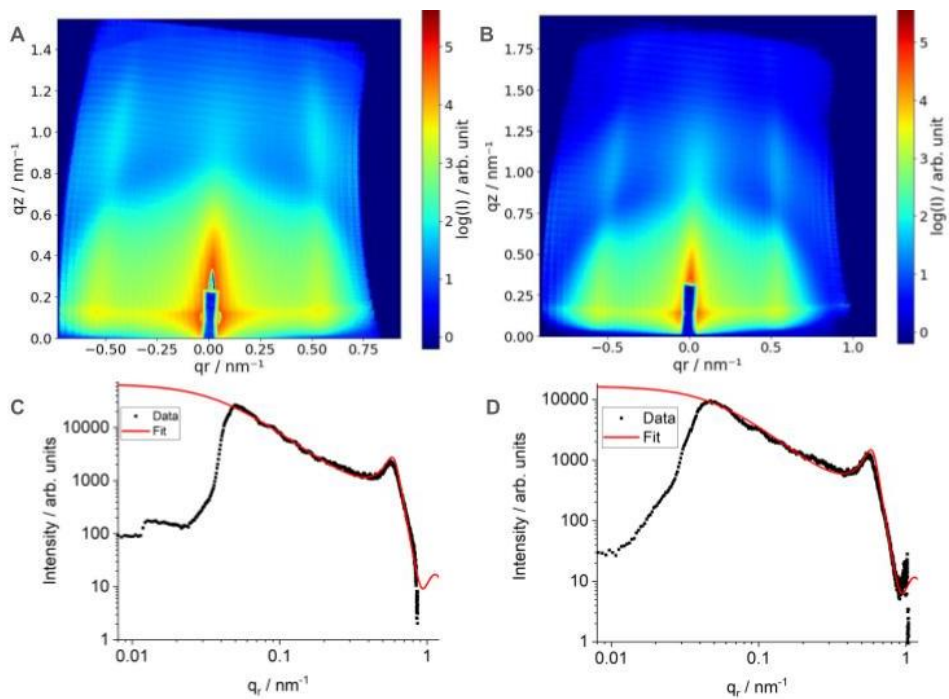

**Figure S2**: Scattering pattern (A, B) and fitted in-plane line (C, D) cuts for NaYF$_4$ particles capped with oleic acid before (left) and during treatment with 0.04 mBar O$_2$ (right).

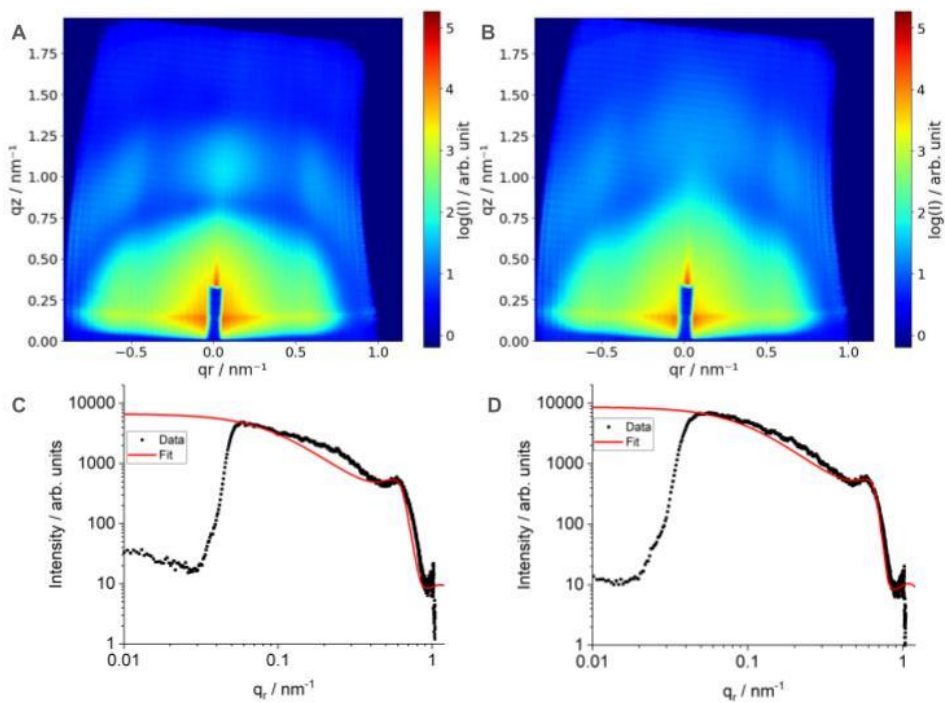

**Figure S3**: Scattering pattern (**A, B**) and fitted in-plane line (**C, D**) cuts for NaYF$_4$ particles capped with oleic acid before (left) and during treatment with 0.13 mBar O$_2$ (right).

For a better comparability of the effects of oxygen treatment on the oleic acid moieties, structure factors have been held constant for before and after samples. The core form factor was restricted to 5 % size changes and no scattering length density changes, while the shell fitted without restrictions. The deviations from fitted curves and the data for all samples at q-values around 0.2 nm$^{-1}$ can be explained by the presence of a variety of more broadened structure factors that are caused by the island-like layer growth of the adsorbed particles and not individually fitted. In Figure S2, the sample before oxygen treatment was measured at a photon energy of 1000 eV instead of the 1240 eV for the other samples. This leads to a $q$ vector that is 1.24 times smaller. The evident comparability of the before and after measurements of these samples under two photon energies shows that the used models and parameters can be transferred to modified conditions.

The ability to track changes of the thickness of the oleic acid moiety for the different samples indicates the robustness of the evaluation method used and the preference for the particles form factor at the chosen cut angle. It also speaks for a nonlocal minimum for the used models, as they converge for all samples. This is despite their clear variation in scattering pattern, due to difference in nanoparticle stacking and island size during as a result of variations in coverage.

*Section 2 – Simulated X-ray scattering patterns*, assessment of sensitivity

The optimal fit between simulated and experimental data yields a core radius of 4.9 nm, a core density of 0.55x the reported density of NaYF$_4$ upconversion nanoparticles, a shell thickness of 1.5 nm and a shell density of 1 x oleic acid. Optical constants for the simulated model were extracted from CXRO Henke tables[1] using the data for Si, SiO$_2$, NaYF$_4$, and oleic acid (C$_{18}$H$_{34}$O$_2$, density =0.895 gcm$^{-3}$)[6] at a photon energy of 1240 eV. The simulated particles are placed in islands with a hexagonal closely packed (FCC) lattice structure and a lattice constant of 12 nm (see Figure S4 B). This leads to the appearance of strong secondary peaks at $q$ = 0.52 nm$^{-1}$, similar to the experimental data. The used model includes an adsorbed layer of oleic acid on the substrate material with a thickness of 4 nm. The adsorbed particles are partly sitting on top of this layer or they are suspended in this layer. This leads to two populations of nanoparticle layers with the higher population being 4.1 nm above the substrate and the lower population being about 1.2 nm above the substrate and polydispersity of both core and ligand shell of 0.12 (see Figure S4C). As the particle densities deviated from the literature data, this model was not utilized to fit the shape and structure of the O$_2$ treated sample as shown in Figure 6 in the main text.

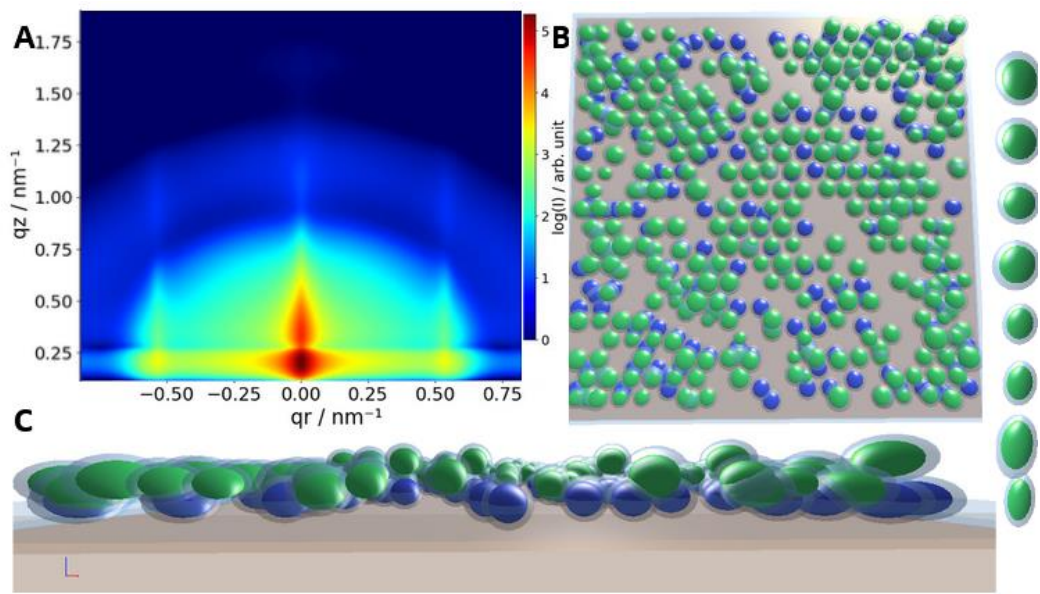

**Figure S3**: Simulated X-ray scattering pattern (**A**) based on the model shown in (**B**) and (**C**).

To circumvent limitations of the definable geometry in BornAgain simulation package, a workaround with three different interparticle distances of 9.8 nm, 11.6 nm and 12.9 nm and substrate-NP heights of 1.5 nm, 0.8 nm and 0.6 nm in the densely packed first layer was used. The partial coverages of the three layers were 10, 10 and 60 %, respectively. For example, the structure factor at higher out-of-plane scattering vectors (vertical rods) is represented more strongly in the experimental data than in the simulation. To minimize this discrepancy, particle agglomerates as observed in the AFM images can be added to the model as paracrystals (as used earlier in the analytical fit).

Figure S5A represents the horizontal and vertical line cuts taken from the simulation shown in Figure S4. Figures S5B and S6-S8 demonstrate the sensitivity of different parameter variations on resulting fit. Deviations of the simulated and measured data at high $q$ values can be explained by a background radiation coming from Compton scattering that was not added to the simulation to maintain good contrast for the higher order minima in simulated scattering pattern. The deviation at $q$ lower than the first order minimum in the out-of-plane scan is not yet fully understood but might be rooted in the geometry of the used scattering setup, as they are apparent in other datasets collected at the APPEXS endstation at beamline 11.0.2 at the ALS.

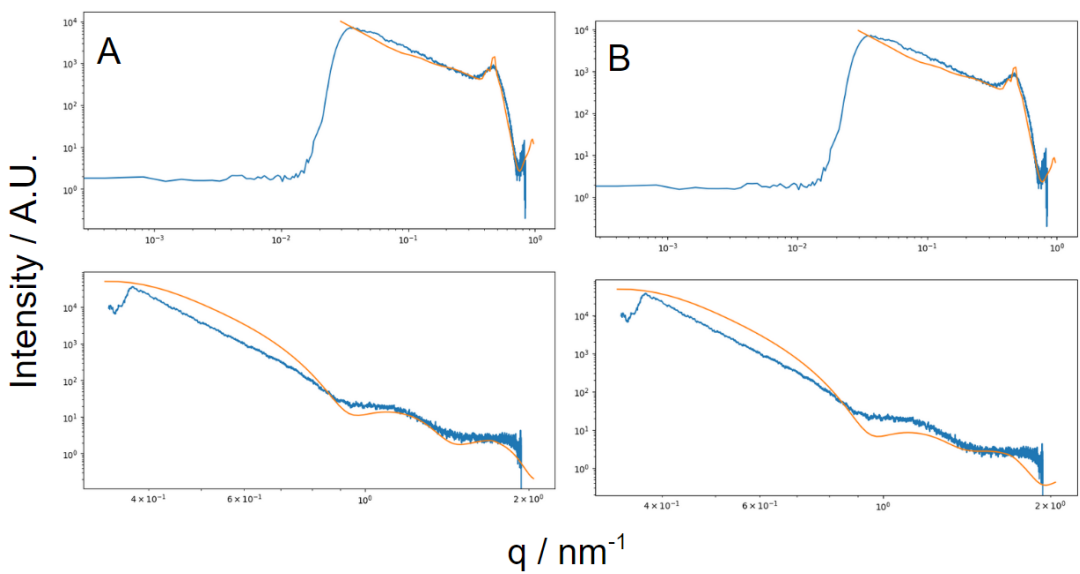

**Figure S5:** In-plane (top) and out-of-plane line cuts (bottom) taken from the measured scattering pattern before $O_2$ treatment (blue) and from simulated data (orange). **A** Core diameter of 4.9 nm, a core density of 0.55x the reported density of NaYF4 upconversion nanoparticles, a shell thickness of 1.5 nm and a shell density of 1x oleic acid. **B** Core radius is reduced by 0.4 nm to a total core radius of 4.5 nm.

As shown in Figure S5 B upon decreasing the size of the NaYF$_4$ nanoparticle by 0.4 nm, while maintaining an overall radius of 6.4 nm leads to significant changes in both in-plane and out-of-plane scattering curves. The in-plane curve (Figure S5 B top) shows a shift of the form factor related minimum intensity to higher $q$ values and a slight decrease in overall scattering intensity. The out-of-plane curve shows more severe changes as period of the minima is modulated to higher $q$ with changing overall particle density and increased thickness of the oleic acid ligand shell.

Upon increasing the ligand layer thickness from 1.3 to 1.7 nm, the overall scattering intensity is slightly reduced, while the first order minimum is shifted to lower $q$ values (compare Figure S6 A and B). Additionally, the period between the first and second order minimum becomes longer and the amplitude of the second order minimum increases above experimental values.

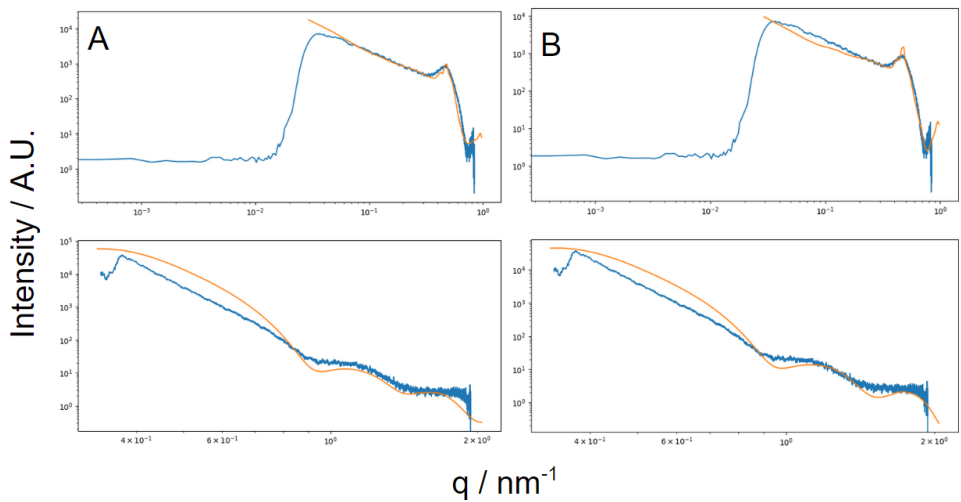

**Figure S6:** In-plane (top) and out-of-plane line cuts (bottom) taken from the measured scattering pattern before $O_2$ treatment (blue) and from simulated data (orange). **A** Shell thickness is increased by 0.2 nm to a total core+shell radius of 6.6 nm. **B** Shell thickness is reduced by 0.2 nm to a total core+shell radius of 6.2 nm.

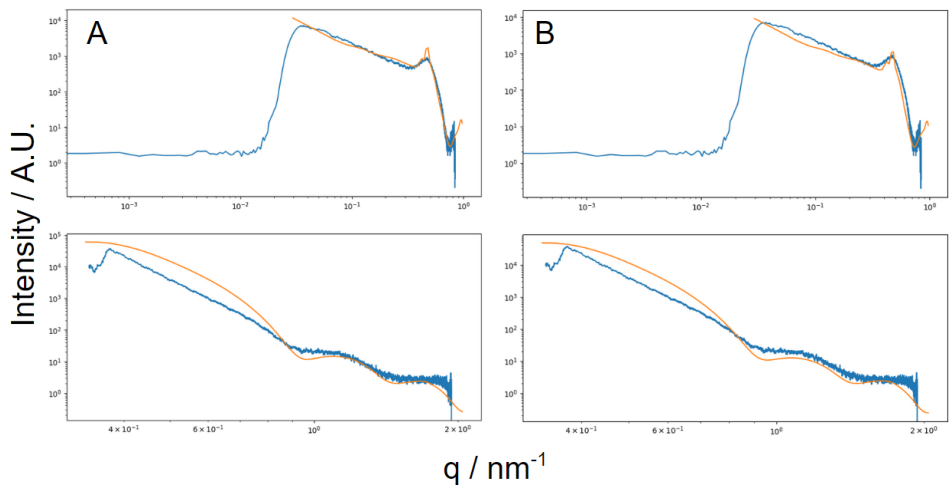

**Figure S7:** Horizontal (top) and vertical line cuts (bottom) taken from the measured scattering pattern before $O_2$ treatment (blue) and from simulated data (orange). **A** Core density is increased by 0.05 to a total density of 0.6 x $NaYF_4$. **B** Core density is reduced by 0.05 to a total density of 0.5 x $NaYF_4$

To follow the dependence of the simulated scattering pattern on the density of the nanoparticle and oleic acid moiety, the optical constants were calculated using the procedure described above, while the density was varied. A variation of the optical constants of the $NaYF_4$ nanoparticles leads to an increase/decrease in the total scattering intensity for increased/decreased values (compare Figure S7 A and B). Additionally, the influence of the ligand moiety on the overall form factor increases with decreasing $NaYF_4$ density due to a decreased contrast between nanoparticle and shell, leading to a shift of the first order maximum to higher $q$ in both in and out-of-plane scans and vice versa for increased density. Similar effects can be observed with a change in the shell density, as the first order minimum shifts to higher $q$. Additionally, the amplitude of the first order minimum increases due to an increased contrast between particle and ligand shell, while the amplitude of the second order minimum, associated with the overall particle size decreases, as the contrast between vacuum and the ligand decreases (see Figure S8 A). A reduction of the polydispersity for the core/shell form factor leads to more pronounced minima (see Figure S8 B).

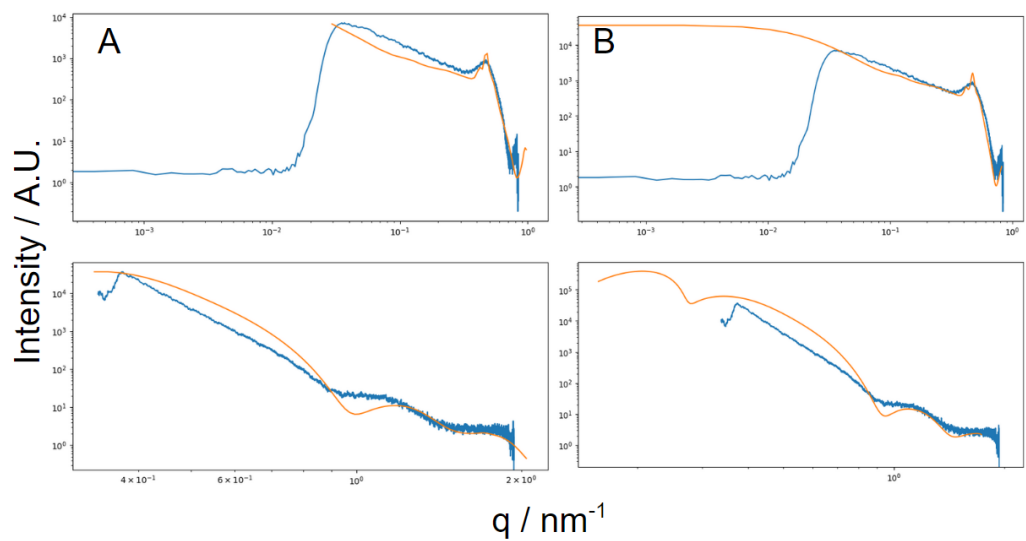

**Figure S8:** Horizontal (top) and vertical line cuts (bottom) taken from the measured scattering pattern before $O_2$ treatment (blue) and from simulated data (orange). **A** Shell density is reduced by 0.5 to a total density of 0.5x $NaYF_4$. **B** Core/shell polydispersity is reduced from 0.28 to 0.25

. By comparing the variations of the particle input parameter with their visible influence on the scattering curve we can conclude, within the constraints of the used model, an error of the evaluated particle radius, shell thickness, particle and shell thickness and polydispersity below the demonstrated values. In values, the evaluated size is 4.9 nm ± 0.1 nm, the evaluated thickness 1.6 ± 0.1 nm, the evaluated particle density 3.07 ± 0.14 gcm$^{-3}$, the shell density 0.895 ± 0.224 gcm$^{-3}$ and the polydispersity 0.28 ± 0.015.

*Section 3 – Simulated X-ray scattering pattern after $O_2$ exposure*

To circumvent limitations of the definable geometry in BornAgain simulation package, a workaround with three different interparticle distances of 9.8 nm, 11.6 nm and 12.9 nm and substrate-NP heights of 1.5 nm, 0.8 nm and 0.6 nm in the densely packed first layer was used. The partial coverages of the three layers were 10, 10 and 60 %, respectively. For example, the structure factor at higher out-of-plane scattering vectors (vertical rods) is represented more strongly in the experimental data than in the simulation. To minimize this discrepancy, particle agglomerates as observed in the AFM images can be added to the model as paracrystals (as used earlier in the Section 1 - analytical fitting).

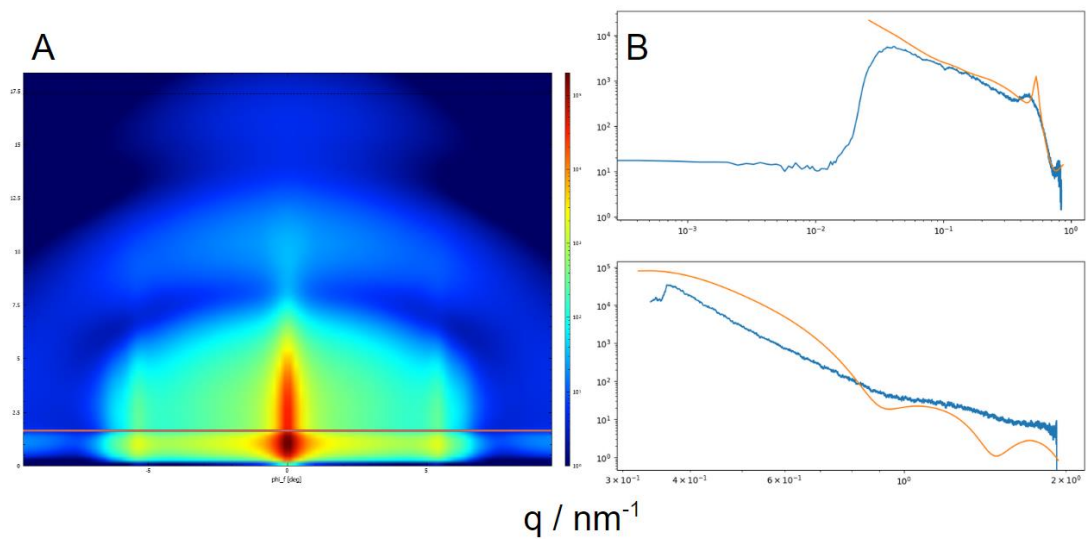

**Figure S9**: Simulation of oleic acid nanoparticles after $O_2$ treatment, with focus on maintaining the strong secondary peaks at 5.5 degrees associated with the structure factor. A Simulated scattering pattern. B Comparison between simulated (orange) and measured (blue) line cuts along the horizontal axis (top) and the vertical axis (bottom).

Figure S9 shows a simulated scattering pattern for the oleic acid capped $NaYF_4$ nanoparticle system after a treatment with 0.007 mBar $O_2$ under derived from the same model as shown in Figure S4. The simulation was modified by reducing shell thickness to 0.8 nm, increasing shell density 1.4 times, removing the oleic acid layer on the substrate and decreasing the number of nanoparticles. The big central feature is given by patches of oleic acid on the substrate with a height of 5 nm and an average radius of 15 nm, with a polydispersity of 0.5. The scattering pattern of this model represents the structure factor derived peaks at $q=0.52$ nm$^{-1}$ visually better than the model in the main text (see Figure S9A). Additionally, the appearance of the second order maximum exhibits more similarities to the recorded data. Despite this, the simulated data does not agree quantitatively both in out-of-plane and in-plane cuts, with the highest deviations in the amplitude of the first and second order minima, and the shape and amplitude of the structure factor maximum.

*Section 4 – Electron density reconstruction of oleic acid capped NaYF4 nanoparticles*

To verify the assumption of the simulated model in figure 7 of the main text, we conducted an independent analysis using an electron density reconstruction of the out-of-plane scattering curve of the $O_2$ treated sample, shown in figure S5A. To generate the reconstruction, pair distance distribution function (PDDF) of the curve was performed using BioXTAS RAW 2.[7] The used PDDF assumed a monodisperse arbitrary shape, a maximum particle size of 11.4 nm, and the first 200 points were truncated for the fit. A quality of fit value of 71 was reached for the PDDF. Electron density reconstruction was performed using the DENSS code and a total of 200 electrons.[8] The reconstruction yielded a radius of gyration of 4.24 nm and a mean RSC of 0.99 with a low ambiguity of the reconstructed model. The electron density reconstruction was drawn in a volume representation with PYMOL[9] as shown in Figure S10.

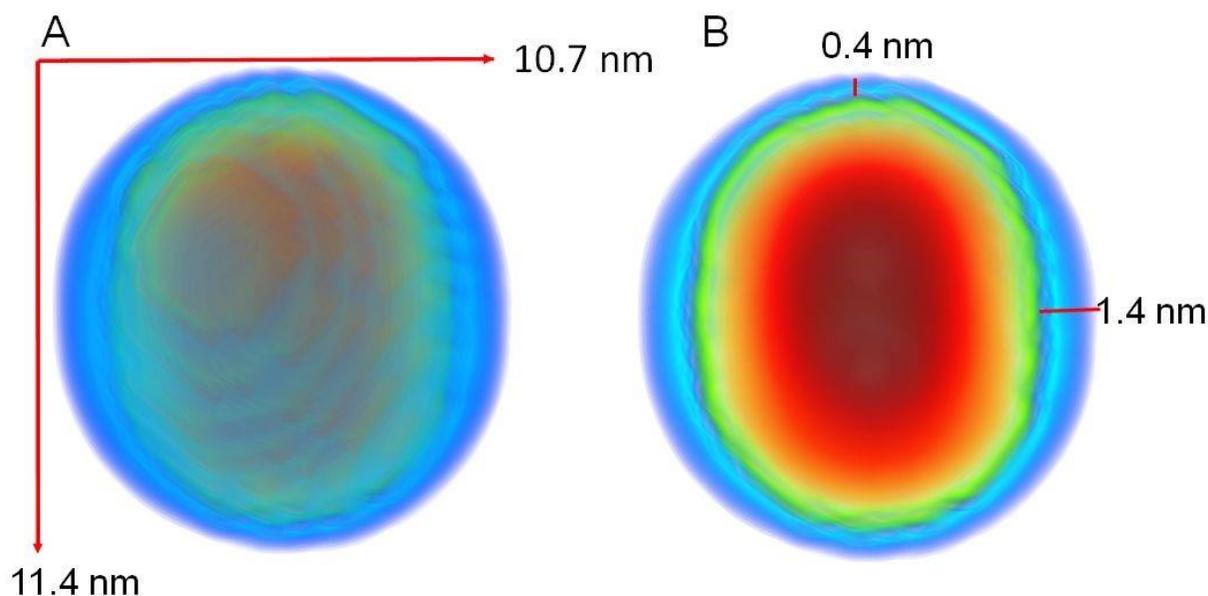

**Figure S10**: Electron density reconstruction of the out-of-plane scan of 9 nm $NaYF_4$ nanoparticles capped with 1.6 nm oleic acid after treatment with 5 mTorr $O_2$. The color map marks electron density with dark red being more than 12 electrons/voxel, light red between 7 and 12, yellow between 7 and 4, turquoise between 4 and 1, and dark blue below 1. **A** Volume representation with scale bare derived from the radius of gyration of the particle. **B** Slice representation of the particle center with indicators for the thickness of the oleic acid layer.

The electron density reconstruction reveals an elongated nanoparticle with a higher shell thickness on the equatorial axes when compared to the longitudinal axes (see Figure S10 B). This structure suggests the two main assumptions of the simulation model shown in the main text.

First, with the knowledge that the NaYF4 nanoparticles are nearly perfectly spherical, derived from AFM, TEM and SEM the elongated shape implied different distances of the particle center from the substrate as the interaction between reflectivity and form factor decreases the $q$ value of the first order minimum for particles sitting further away from the surface, leading to an apparent increase in particle size in the out-of-plane direction.

Second, the shell in the elongated direction is thinner, this implies a higher in-plane thickness and therefore an inhomogeneous etching of the ligand moiety under $O_2$ influence, with the highest rates directly on top of the particles. As the DENSS reconstruction is an ab initio method and not fed with any assumption from the simulation, this similarity in results implies that the used model and drawn conclusions reflect the correct geometry with a robust accuracy.

*Section 5 – Additional information from C 1s APXPS spectra*

In the main manuscript, the evolution of the functional groups ($CH_x$, COH, C=O, COOH) as a function of oxidation time is shown for the four different $O_2$ pressures (Figure 2). For completeness, we show in the following additional information that can be extracted from C 1s spectra (such as those shown in Figure 1 of the main manuscript).

Figure S9a displays the total carbon amount for the four different experiments as a function of reaction time. This value was calculated by adding up the four C 1s components (CHx, COH, C=O, COOH). The data of each pressure series are normalized to unity for a data point taken at the beginning of the oxidation reaction, after about 4 min. Fig. S9a shows a decrease in the amount of carbon with reaction time, in agreement with a gradual volatilization of the ligand layer around the NPs.

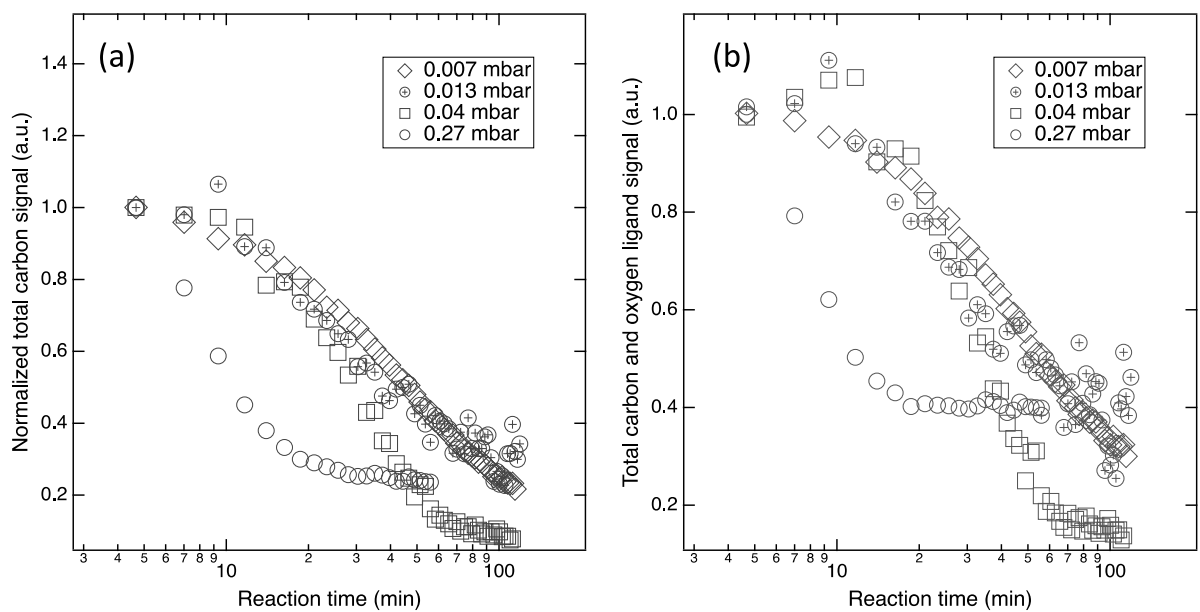

**Figure S9.** **(a)** Total C 1s signal as a function of reaction time for the four different $O_2$ reaction pressures. **(b)** Total carbon and oxygen content deduced from the C 1s signal, under consideration of the C/O stoichiometry of the different C species.

The data in Fig. S9a are not sufficient to judge the change in the thickness of the ligand layer with reaction time. For this, the addition of oxygen due to the functionalization in the oxidation process needs to be taken into account. The total amount of material in the ligand layer (neglecting hydrogens) is then calculated from the C 1s peak components areas according to (1*CHx)+(2*COH)+(2*C=O)+(3*COOH), where the factors reflect the abundance of C and O in each functional group. The data of each pressure series are then again normalized to unity for a data point taken at the beginning of the oxidation reaction, after about 4 min, just like in the case of the total C data in Figure S9a.

The results are shown in Figure S9b. For the reaction at 0.013 mbar and 0.04 mbar $O_2$ a slight increase in the total amount of C+O is observed at the beginning of the oxidation, but this increase is in the range of the experimental error. A clear increase in the carbonaceous layer, as previously observed for the oxidation of coronene[10] is thus not present for the oxidation of the oleic acid layer, where volatilization of the ligand layer dominates over the addition of material through functionalization.

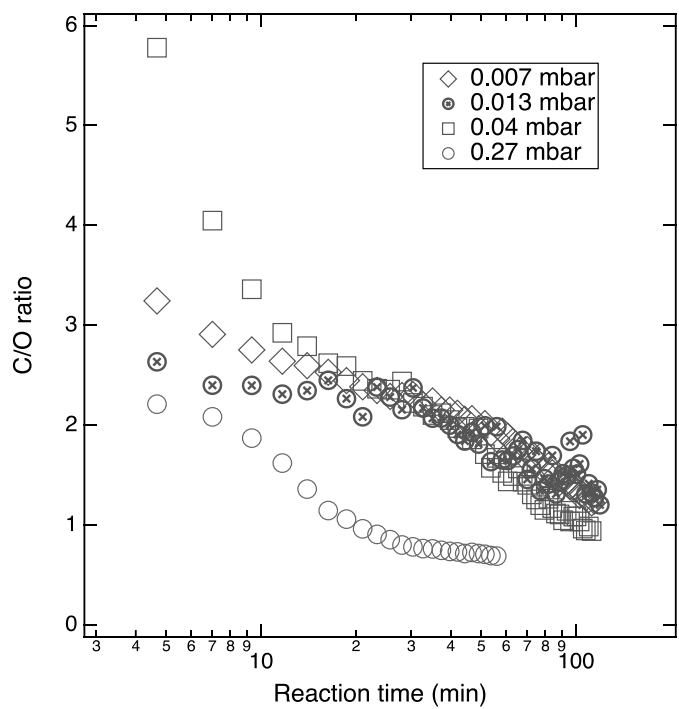

**Figure S10.** C/O ratio as a function of reaction time, deduced from the C 1s signal.

Another metric that can be extracted from the C 1s data is the C-to-O ratio as a function of reaction time, which informs on the degree of functionalization. This ratio is shown in Fig. S10 for the four different oxidation experiments. For the experiment at the highest $O_2$ pressure (0.27 mbar), where the reaction proceeds fastest, a C-to-O ratio close to unity is reached already after 30 min reaction time, while for the other $O_2$ pressures this value is approached but not quite reached within the duration of the oxidation of about 120 min. A C-to-O ratio close to unity is consistent with an average final reaction product similar to acetic acid.

# References

[1] https://henke.lbl.gov/optical_constants/getdb2.html

[2] Mackenzie, L.E., Goode, J.A., Vakurov, A. et al., The theoretical molecular weight of $NaYF_4$:RE upconversion nanoparticles, *Sci Rep.* 8, 1106 (2018).

[3] A. Guinier and G Fournet, *Small-Angle Scattering of X-Rays*, John Wiley and Sons, New York, (1955)

[4] J. Teixeira, *J. Appl. Cryst.*, 21, 781-785 (1988).

[5] H. Matsuoka et. al. *Physical Review B*, 36 (1987) 1754-1765; Hideki Matsuoka et. al. *Physical Review B*, 41, 3854-3856 (1990).

[6] https://pubchem.ncbi.nlm.nih.gov/compound/Oleic-Acid

[7] J. B. Hopkins, BioXTAS RAW 2: new developments for a free open-source program for small-angle scattering data reduction and analysis., *Journal of Applied Crystallography* 57, 194-208 (2024).

[8] T. D. Grant, Ab initio electron density determination directly from solution scattering data, *Nature Methods* 15, 191–193 (2018).

[9] The PyMOL Molecular Graphics System, Version 3.0 Schrödinger, LLC.

[10] E.R. Mysak, J.D. Smith, J.T. Newberg, P.D. Ashby, K.R. Wilson, H. Bluhm, Competitive Reaction Pathways for Functionalization and Volatilization in the Heterogeneous Oxidation of Coronene Thin Films by Hydroxyl Radicals and Ozone, *Phys. Chem. Chem. Phys.* 13, 7554 (2011).